\documentclass[sigconf,dvipsnames,svgnames,screen]{acmart}

\usepackage[utf8]{inputenc}

\usepackage[english]{babel}

\usepackage{graphicx}
\usepackage{subcaption}
\usepackage{longtable}
\usepackage{hyperref}
\usepackage{url}
\usepackage{xspace}
\usepackage{xcolor}
\usepackage{amsfonts,amsmath}
\usepackage{enumitem}
\usepackage{booktabs}
\usepackage{cleveref}               
\usepackage{multicol}
\usepackage[abbreviations]{foreign} 
\usepackage{tikz}
\usetikzlibrary{arrows.meta,decorations.pathreplacing,patterns}
\usepackage{tcolorbox}
\usepackage{svg}
\usepackage{multirow}
\usepackage{colortbl}
\usepackage{arydshln}

\definecolor{pastelgreen}{HTML}{D3E6D6}
\definecolor{pastelviolet}{HTML}{DDD3EC}
\definecolor{hatchcolor}{HTML}{AFA8C4}
\definecolor{softgreen}{HTML}{79B482}
\definecolor{softviolet}{HTML}{8F7BB3}
 
\newcommand{\SumupBoxViolet}[2][]{
    \begin{tcolorbox}[colbacktitle=softviolet!30, colframe=softviolet!30,colback=pastelviolet!30, coltitle=black,title=\textbf{#1}, left=3pt, right=3pt, top=2pt, bottom=0pt]
        #2
    \end{tcolorbox}
}

\newcommand{\rennes}{Rennes\xspace}
\newcommand{\nancy}{Nancy\xspace}

\newcommand{\cfg}[1]{\texttt{#1}\xspace}
\newcommand{\defconfig}{\cfg{defconfig}}
\newcommand{\allmodconfig}{\cfg{allmodconfig}}
\newcommand{\allyesconfig}{\cfg{allyesconfig}}

\AtBeginDocument{%

}

\title{Trading Joules for Options: Exploring The Energy Price of Building Linux Kernel Configurations}

\author{Sophie Fortz}
\orcid{0000-0001-9687-8587}
\email{sophie.fortz@inria.fr}
\affiliation{
  \institution{Inria, Univ. Rennes, CNRS, IRISA}
  \city{Rennes}
  \country{France}
}
\author{Quentin Perez}
\orcid{0000-0002-1534-4821}
\email{quentin.perez@irisa.fr}
\affiliation{
  \institution{INSA Rennes, Univ. Rennes, CNRS, Inria, IRISA}
  \city{Rennes}
  \country{France}
}
\author{Mathieu Acher}
\orcid{0000-0003-1483-3858}
\email{mathieu.acher@irisa.fr}
\affiliation{
  \institution{INSA Rennes, Univ. Rennes, CNRS, Inria, IRISA, IUF} 
  \city{Rennes}
  \country{France}
}

\setcopyright{none}
\renewcommand\footnotetextcopyrightpermission[1]{}
\begin{document}

\begin{abstract}
With more than $20{,}000$ configuration options, the Linux kernel is one of the largest open-source product lines ever built, so the energy spent configuring and building it matters at scale. Measuring that energy is challenging because results can change with the machine used, the metric observed, and the number of repetitions.

We present an empirical framework to measure Linux kernel build energy on bare-metal nodes of the Grid’5000 testbed using three complementary metrics: RAPL counters (CPU package and DRAM), the node's BMC, and a rack PDU. We evaluate it on $3{,}750$ builds (15 configurations, 25 runs, 10 machines, 2 clusters) and extend the proxy analysis with 600 additional builds on 200 random configurations (1 run on each of 3 machines).

Our results translate into practical guidelines. First, absolute energy values must be calibrated per machine: even within a homogeneous cluster, measurements differ by stable machine-specific factors (up to $8.0\%$ for RAPL and $9.2\%$ for the PDU). In contrast, comparisons between configurations are robust when the design is balanced across machines. Second, the three metrics agree on the overall picture for builds longer than about one minute: BMC reports on average $7.0\%$ less energy than PDU, and PDU power is well explained by a constant baseline plus a term proportional to RAPL. Third, build time is a good proxy to rank configurations by energy, while kernel size is less reliable. 
\end{abstract}

\ccsdesc[500]{Software and its engineering~Software performance}
\ccsdesc[300]{Software and its engineering~Empirical software validation}
\ccsdesc[300]{Computer systems organization~Power and energy}

\keywords{Energy consumption, Linux kernel, Deep variability, Software configurations, Empirical study}

\maketitle

\section{Introduction}

Data centers consumed approximately 1.5\% of the world's electricity in 2024, and
this consumption is expected to more than double by
2030~\cite{iea2025energyai}. Hardware efficiency gains have so far absorbed much of this growth~\cite{masanet2020recalibrating}, but the energy spent also depends on the software that runs~\cite{pereira2017energy}, and is now
evaluated alongside execution time~\cite{ournani2020taming}. However, to make such software-level assessments actionable, we need energy measurements that are both reliable and comparable.

In practice, reliable energy measurement remains difficult because every instrument involves a trade-off between coverage, sampling frequency, and accuracy~\cite{DavidGHKL10,kavanagh2019rapid,raffin2025dissecting}. Even when the software is unchanged, the measurements fluctuate due to hardware heterogeneity and power-management choices such as C-states or Turbo Boost~\cite{ournani2020taming}. Without a clear protocol (how many runs, how many machines, and which sensor to rely on), an apparent improvement between two variants may simply be measurement noise rather than a true software effect.

This issue is particularly significant for configurable software systems, whose non-functional properties depend on both the configuration and the underlying software and hardware layers~\cite{SiegmundGAK15,MartinTSE2021}. The compilation of the Linux kernel is a good case study: compilation is a canonical compute-intensive benchmark~\cite{bucek2018spec,pts2024buildlinux}, kernel builds are executed at scale in continuous integration environments, and kernel configurations induce build durations ranging from 25\,s to more than 1000\,s, with corresponding variation in power profiles. Nevertheless, previous investigations of Linux configurations focus on binary size and build time~\cite{Acher0LBJKBP22,BorgesPKA25}, whereas energy-oriented studies of configurable systems keep the lower layers fixed~\cite{GuegainQR21,TernavaLPKAC25,WeberKSAS23}.
Studying energy across such a configuration space means measuring at scale. With more than 20,000 options, even a modest sample of configurations represents hundreds of builds, each multiplied by repetitions, machines, and sensors. A careful protocol quickly becomes unaffordable: our calibration campaign alone (15 configurations, 25 runs, 10 machines) took 535 hours of build time.
Before scaling up, we therefore need to establish to what extent the measurement infrastructure can be trusted, which pitfalls to avoid, and where cost can be reduced without losing accuracy. That is, how many runs, on how many machines, with which sensor, and whether a cheap proxy such as build time can be leveraged to actually measure energy.

In this paper, we present a framework that measures the energy of Linux
kernel builds on bare-metal Grid'5000 nodes with RAPL, BMC and PDU, and
evaluate it on 3,750 builds: 15 configurations, built 25 times on each of
five identical machines in two clusters. We study how to calibrate the
framework, whether the instruments agree, and whether build time or kernel
size can replace energy measurement. We find that machines with identical hardware differ by stable factors of up to 9\% in absolute energy, so comparisons between configurations remain valid. Moreover two to four runs are sufficient for RAPL and the PDU, whereas the BMC is unreliable for builds shorter than one minute.
We then use the resulting lightweight protocol (one run on each of three machines) to measure 200 random configurations.
At this scale, build time remains a good proxy to rank configurations by energy ($\tau_b \approx 0.85$--$0.89$) but only a rough estimator of its value, while kernel size is a poor proxy.


The remainder of this paper is organized as follows.
\Cref{sec:background} provides background, and
\Cref{sec:related-work} discusses related work.
\Cref{sec:problem-statement} states our research questions and
\Cref{sec:methodology} describes the experimental setup.
\Cref{sec:results} presents the results,
\Cref{sec:threats} discusses threats to validity, and
\Cref{sec:conclusion} concludes.

\section{Background}
\label{sec:background}
This section introduces the energy measurement methods that we rely on and how Linux kernel variants are configured and built.

\paragraph{\textbf{Measuring the Energy Consumption of Software}}

Software draws on several hardware resources at once (CPU, memory, storage, network), which makes its energy hard to measure.
Hardware power meters, such as those integrated into Power Distribution Units (PDUs) or into the Baseboard Management Controller (BMC) of servers, measure the power of the whole node, but only at a coarse time granularity (one sample every 1 to 5\,s in our setup).
Intel's Running Average Power Limit (RAPL) interface~\cite{DavidGHKL10} complements them with energy counters integrated into the CPU and updated about every millisecond~\cite{JayOLTOF23}. RAPL reports the energy of several power domains, such as the CPU package and the DRAM, through Model-Specific Registers readable from the operating system (OS).
Software-based power meters such as PowerAPI~\cite{BourdonNRS13} and SmartWatts~\cite{FieniRS20} go further and estimate the energy of a process or container~\cite{perez2024software}. However, they require Linux control groups~\cite{Rub1}, which our infrastructure does not expose, and they infer power from resource-usage metrics rather than measuring it, which adds assumptions and calibration steps.
We therefore rely only on measured power: RAPL, BMCs and PDUs.

\paragraph{\textbf{Configuring and Building Linux Kernel Variants}}

The Linux kernel has more than 20,000 options, described in Kconfig files (build configuration files). Most are boolean (\texttt{y}, \texttt{n}) or tristate (\texttt{y}, \texttt{n}, or \texttt{m} to build the option as a loadable module), and constraints between options make many combinations invalid.
Valid configurations are therefore obtained from predefined targets (\texttt{tinyconfig}, \defconfig, \texttt{allnoconfig}, \allmodconfig, \allyesconfig) or from \texttt{randconfig}, which generates random valid configurations.
A kernel is built with Kbuild, the Make-based build system of Linux~\cite{kernel_kbuild}, which produces the kernel image and its modules. The result also depends on the target architecture, the compiler, and the tools installed on the machine. We thus define a \emph{build} as one execution of Kbuild for a given kernel revision, configuration, architecture, toolchain, and build target.
To keep the toolchain identical across machines, we run Kbuild through TuxMake\footnote{\url{https://tuxmake.org/}}, which executes each build in a versioned container image and makes it reproducible from its command line.
\section{Related Work}
\label{sec:related-work}

Software variability has been extensively studied through the lens of exploration space and its quantification, constraints, and associated non-functional properties. By contrast, research connecting variability to energy consumption remains scarce and limited to software product lines (SPLs).

\paragraph{\textbf{Software Variability and Non-Functional Properties}}

The impact of configuration choices on non-functional properties has been extensively studied~\cite{PereiraAMJBV21}. Performance-influence models capture the effects of options and their interactions~\cite{SiegmundGAK15,SiegmundKKABRS12}, and can be inferred from limited configuration samples~\cite{GuoCASW13} or integrated into Kconfig-based SPLs while requiring only minor adjustments to the build pipeline~\cite{FrieselEK0S22}. However, these models rarely generalize across different layers of variability: their predictions depend on workloads~\cite{MuhlbauerSKDAS23b} and compile-time options used to build the system~\cite{LesoilATBJ21}. In the context of the Linux kernel, most quantitative studies focus on binary size. \citeauthor{Acher0LBJKBP22}~\cite{Acher0LBJKBP22} learned an accurate and interpretable size model from 95,854 configurations using feature selection, and \citeauthor{MartinTSE2021}~\cite{MartinTSE2021} reported that these models degraded between kernel versions, then proposed a transfer-learning solution. \citeauthor{BorgesPKA25}~\cite{BorgesPKA25} published a dataset with over 240,000 configurations annotated with binary size and compilation time. Lastly, although performance often correlates with energy consumption across configurations, this relationship may not hold for all configurations~\cite{WeberKSAS23}. Yet, no study quantifies the energy consumed when building configurations or investigates the influence of the measurement setup.

\paragraph{\textbf{Software Variability and Energy Consumption}}

Energy consumption has only recently been investigated in the context of SPLs. 
\citeauthor{Couto2017}~\cite{Couto2017} proposed a static technique to estimate the worst-case energy consumption across products, without generating and measuring them individually.
\citeauthor{GuegainQR21}~\cite{GuegainQR21} inferred the energy impact of features and their interactions from a sample of products to derive lower-consuming products. Both examine how energy consumption varies across configurations at a high software level, namely features, feature interactions, and products. Along the same line of controlling variability, \citeauthor{TernavaLPKAC25}~\cite{TernavaLPKAC25} study feature reduction as a means to reduce feature space of Linux utilities, considering both built-in reduced implementations (ToyBox/BusyBox vs. GNU) and on-demand debloating techniques. They found that reducing variability (fewer features/options, smaller binaries) does not reliably yield lower energy consumption, while execution time remains a strong proxy for energy.
However, these studies abstract lower software and hardware layers through simplified energy models~\cite{Couto2017} or by keeping them fixed~\cite{GuegainQR21,TernavaLPKAC25}. We complement them by explicitly studying how these lower-level factors influence energy variability.

\section{Problem Statement and Research Questions}
\label{sec:problem-statement}

\begin{figure}[t]
    \centering
    \includegraphics[width=1\linewidth]{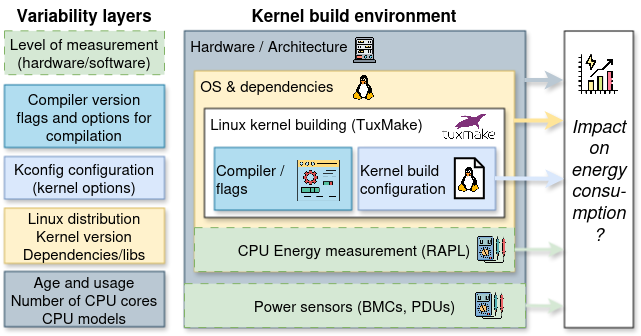}
    \caption{Linux kernel building's deep variability.}
    \label{fig:variability-stack}
    \Description{Linux kernel building's deep variability.}
\end{figure}

The compilation of the Linux kernel constitutes a representative example of \emph{deep software variability}. In line with \citeauthor{LesoilABJ21}~\cite{LesoilABJ21}, we hypothesize that software behavior and non-functional properties are influenced by the interaction of multiple variability layers (variability, including kernel options configuration, OS and its dependencies, hardware).
\Cref{fig:variability-stack} shows this stack of cross-layer variability in our context.

We operationalize this cross-layer variability as a set of controllable factors that can affect the energy and time of Linux kernel builds. At the kernel building layer (``Linux kernel building'' in \Cref{fig:variability-stack}), we vary Kconfig options, affecting the set of compiled translation units (built-in vs. modules) and thus the amount of work performed by the toolchain.
At the OS layer (``OS and dependencies'' in \Cref{fig:variability-stack}), we consider the host OS and the (containerized) build environment, including compiler/toolchain versions and build dependencies. At the hardware layer (``Hardware / Architecture'' in \Cref{fig:variability-stack}), we account for machine characteristics (CPU, number of cores, memory capacity) and dynamic effects such as thermal behavior.

In addition, we explicitly model a measurement layer (``Power sensors'' and ``CPU Energy Measurement'' in \Cref{fig:variability-stack}): we compare energy observations obtained from different instruments (on-chip RAPL counters vs. external BMC/PDU monitoring), whose granularity and sampling rates may yield different estimates over the same build interval.


Based on this problem statement, we formulate three research questions:

\noindent\textbf{$\bullet$ RQ1 — Framework Calibration and Experimental Reproducibility} \emph{To what extent does the proposed measurement framework yield stable and reproducible results across heterogeneous hardware environments?}

\begin{itemize}
\item \textbf{RQ1.1 — Cross-machine and cross-cluster interchangeability.} Are energy measurements statistically equivalent across machines sharing identical hardware specifications? 
We hypothesize that machines within the same cluster yield equivalent measurements up to a constant scaling factor, enabling extrapolation from a subset of machines.

\item \textbf{RQ1.2 — Minimum number of runs for statistical significance.} How many runs are needed for statistically reliable energy measurements? We performed 25 runs per configuration by default (3,750 runs across 15 kernel configurations and 10 machines), without empirically justifying this choice. We seek to identify the minimum run count that preserves statistical validity and reduces experimental costs.
\end{itemize}

\noindent\textbf{$\bullet$ RQ2 — Cross-Sensor Agreement Between RAPL, BMC, and PDU Power Metrics} \emph{To what degree do RAPL, BMC, and PDU measurements agree, and how do component-level and system-level energy estimates relate?}

BMC and PDU capture system-wide power at fixed sampling intervals, while RAPL measures cumulative CPU and DRAM energy over the build. We expect BMC and PDU power profiles to agree and aggregate RAPL readings (CPU + DRAM) to represent a consistent fraction of system-level energy.

\noindent\textbf{$\bullet$ RQ3 — Build Duration and Kernel Size as Proxies for Energy Consumption} \emph{Can build time and compiled kernel size serve as reliable, low-overhead proxies for predicting Linux kernel build energy?}

Energy measurement requires dedicated instrumentation. We investigate whether build time and/or kernel size correlate consistently with energy across configurations, enabling estimation where direct measurements are unavailable.

\section{Methodology}
\label{sec:methodology}

\begin{figure*}[t]
    \centering
    \includegraphics[width=\linewidth]{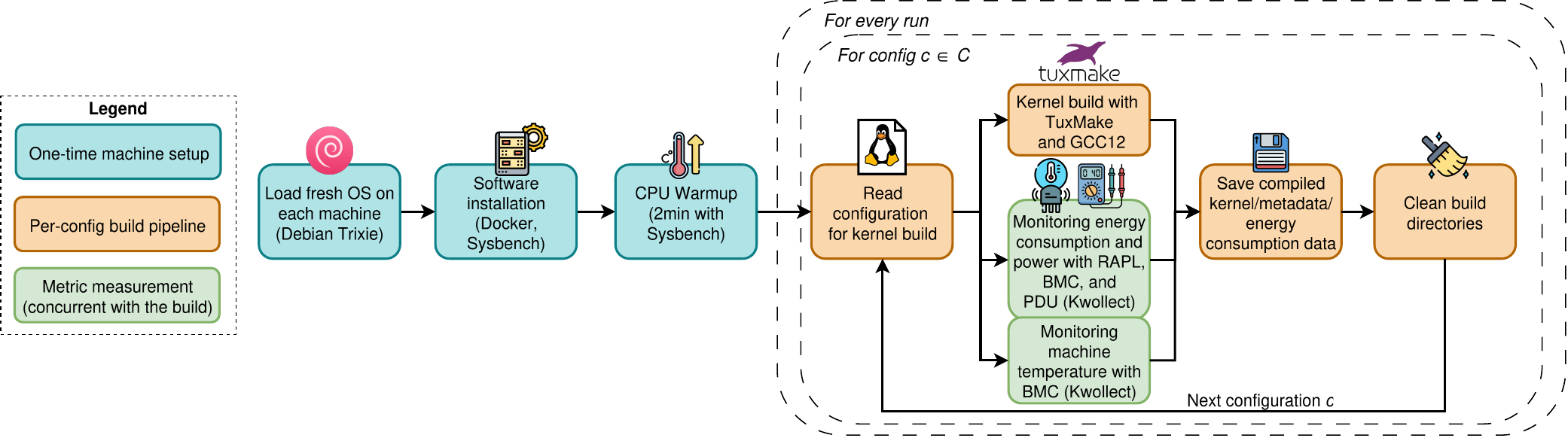}
    \caption{Overview of the experimental methodology. After a one-time setup of each machine, every selected kernel configuration is repeatedly built with TuxMake and GCC 12 following a 2-minute CPU warm-up, while energy (RAPL, BMC, PDU) and temperature (BMC) are monitored concurrently via the Kwollect interface.}
    \Description{Overview of the experimental methodology.}
    \label{fig:confvert-methodo}
\end{figure*}

Our evaluation is organized in two phases: a first phase covering RQ1--RQ3 on a curated set of 15 configurations across 10 machines, and a second phase extending RQ3 to 200 random configurations. Both rely on the infrastructure illustrated in \Cref{fig:confvert-methodo}. We measure the energy consumed when building Linux kernel~7.0\footnote{\url{https://github.com/torvalds/linux/tree/v7.0}} with TuxMake and GCC~12. \Cref{tab:campaign_cost} summarizes the cost of both phases in terms of build count, cumulative build time, and BMC-measured energy. 

\begin{table}[t]
  \centering
  \small
  \caption{Cost of the experimental campaigns: number of builds, total build time and BMC energy.}
  \label{tab:campaign_cost}
  \begin{tabular}{crrr}
    \toprule
    Experiment & Builds & Build time (h) & BMC energy (J) \\
    \midrule
    15 configurations & \multirow{2}*{3,750} & \multirow{2}*{535.4} & \multirow{2}*{368,594,571} \\
    (25 runs $\times$ 10 machines) &&&\\
    200 configurations & \multirow{2}*{600} & \multirow{2}*{37.3} &  \multirow{2}*{22,229,352} \\
    (1 run $\times$ 3 machines) &&&\\
    \midrule
    Total & 4,350 & 572.8 & \begin{tabular}[t]{@{}c@{}}390,823,922\\($\approx 108.6$\,kWh)\end{tabular} \\
    \bottomrule
\end{tabular}
\end{table}

\subsection{Experimental Infrastructure}
\label{subsec:infrastructure}

All experiments were conducted on Grid'5000 infrastructure~\cite{Bolze06G5k}, a large-scale French experimental platform that gives users exclusive bare-metal access to nodes (machines) and allows them to deploy their own system image. We used two homogeneous clusters located at two different sites: \texttt{parasilo} (\rennes (C1)) and \texttt{gros} (\nancy (C2)), selecting five nodes from each. \Cref{tab:hardware-compact} summarizes their hardware and software configuration.
Within each cluster, we verified both manually and programmatically that all selected nodes were strictly identical. Using two clusters with different CPU generations and memory configurations allows us to assess whether our observations generalize across hardware. All nodes ran Debian~13 (``Trixie'') with Linux kernel 6.12.111. Each node was freshly deployed before every batch of experiments so that all runs start from the same software state. Kernel builds were performed inside a container based on the \texttt{TuxMake GCC~12} image, keeping the toolchain constant across nodes and clusters.

\begin{table*}[htpb]
\setlength{\tabcolsep}{2pt}
\centering
\small
\caption{Hardware and software configuration of the machines used in the experiments (identical within each cluster).}
\label{tab:hardware-compact}
\begin{tabular}{llllllll}
\toprule
\textbf{Cluster} & \textbf{Nodes (Machines)} & \textbf{Server} & \textbf{CPU} & \textbf{RAM (GiB)} & \textbf{Kernel} & \textbf{OS} & \textbf{Storage} \\
\midrule
\rennes (C1) & parasilo-18, 19, 20, 21, 22 & PowerEdge R630 & 2 x Intel Xeon E5-2630 v3 & 128 & 6.12.111 & Debian 13 & 600 GB \\
\midrule
\nancy (C2) & gros-100, 101, 102, 104, 105 & PowerEdge R640 & 1 x Intel Xeon Gold 5220 & 96 & 6.12.111 & Debian 13 & 480 GB \\
\bottomrule
\end{tabular}
\end{table*}

\begin{figure}[htbp]
    \centering
    \resizebox{1\columnwidth}{!}{%
\begin{tikzpicture}[>=Stealth, font=\LARGE,]
 
  \def\xc{5.5}
  \def\tone{3.0}
  \def\ttwo{8.0}
  \def\yload{3.1}   
 
  \draw[->, thick] (0.5,-0.3) -- (0.5,3.6)
    node[pos=0.6, above, sloped] {Power (W)};
 
  \filldraw[fill=pastelgreen, draw=black!70] (0.5,0) rectangle (10.5,0.7);
 
  \filldraw[fill=pastelviolet, draw=black!70] (\tone,0.7) rectangle (\ttwo,\yload);
 
  \fill[pattern=north east lines, pattern color=hatchcolor]
    (\tone,0) rectangle (\ttwo,\yload);
  \draw[thick, black!70] (\tone,0) rectangle (\ttwo,\yload);
 
  \node[fill=white, fill opacity=0.9, text opacity=1, inner sep=2pt]
    at (\xc,0.35) {Node power baseline};
  \node[fill=white, fill opacity=0.9, text opacity=1, inner sep=3pt, align=center]
    at (\xc,{(0.7+\yload)/2}) {Power induced by the load\\(kernel compilation)};
 
  \draw[->, thick] (0.2,0) -- (11.2,0) node[right] {Time (s)};
 
  \draw[decorate, decoration={brace, amplitude=6pt}, thick]
    (\tone,\yload+0.15) -- (\ttwo,\yload+0.15);
  \coordinate (bmid) at ({(\tone+\ttwo)/2},{\yload+0.15});
 
  \node[anchor=south, inner sep=2pt, align=left] at ([yshift=12pt]bmid) {%
    $\displaystyle E_{\mathrm{RAPL}} = E_{\mathrm{RAPL}}(t_2) - E_{\mathrm{RAPL}}(t_1)$\\[8pt]
    $\displaystyle E_{\mathrm{BMC}} = \int_{t_1}^{t_2} P_{\mathrm{BMC}}(t)\,\mathrm{d}t
      \approx \sum_{t_k \in [t_1,t_2]} P_{\mathrm{BMC}}(t_k)\,\Delta t_{\mathrm{BMC}}$,
      \ $\Delta t_{\mathrm{BMC}} = 5\,\mathrm{s}$\\[8pt]
   $\displaystyle E_{\mathrm{PDU}} = \int_{t_1}^{t_2} P_{\mathrm{PDU}}(t)\,\mathrm{d}t
      \approx \sum_{t_k \in [t_1,t_2]} P_{\mathrm{PDU}}(t_k)\,\Delta t_{\mathrm{PDU}}$,
      \ $\Delta t_{\mathrm{PDU}} = 1\,\mathrm{s}$};
 
  \def\ypdu{-1.5}
  \draw[->, thick] (0.2,\ypdu) -- (11.2,\ypdu) node[right] {Time (s)};
  \node[anchor=east, align=center] at (0.1,\ypdu) {PDU\\(1\,Hz)};
  \foreach \x in {0.5,0.75,...,10.5} {\draw (\x,\ypdu-0.12) -- (\x,\ypdu+0.12);}
 
  \def\ybmc{-2.6}
  \draw[->, thick] (0.2,\ybmc) -- (11.2,\ybmc) node[right] {Time (s)};
  \node[anchor=east, align=center] at (0.1,\ybmc) {BMC\\(0.2\,Hz)};
  \foreach \x in {0.5,1.75,...,10.5} {\draw (\x,\ybmc-0.12) -- (\x,\ybmc+0.12);}
 
  \draw[dashed] (\tone,0) -- (\tone,-3.2);
  \draw[dashed] (\ttwo,0) -- (\ttwo,-3.2);
  \node[below, fill=white, inner sep=1pt] at (\tone,-3.2) {$t_1$};
  \node[below, fill=white, inner sep=1pt] at (\ttwo,-3.2) {$t_2$};
 
 
\end{tikzpicture}
}
\caption{Energy of a kernel compilation between $\mathbf{t_1}$ and $\mathbf{t_2}$ (hatched area), measured with RAPL, BMC and PDU.}
\label{fig:illustration-energy-formula}
\Description{Energy of a kernel compilation between [t_1,t_2] (hatched area), measured with RAPL, BMC and PDU.}
\end{figure}
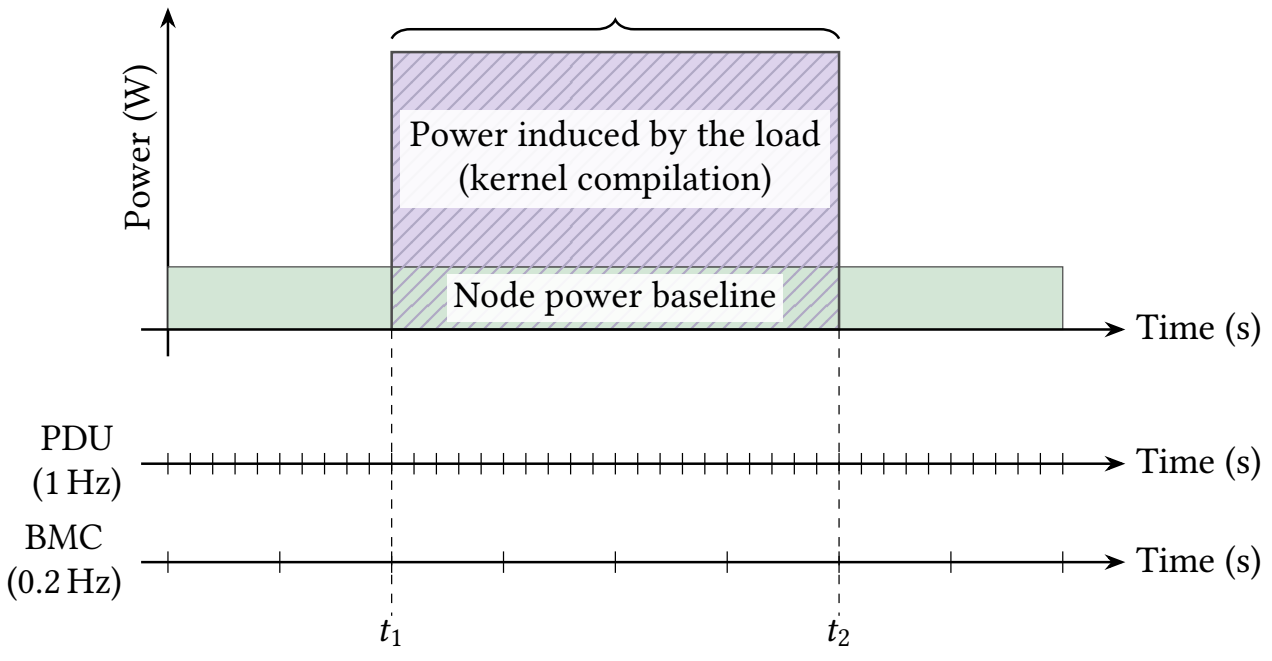

Each cluster is equipped with hardware-based power measurement devices attached to each machine. BMC measurements are available on both \rennes and \nancy, while PDU measurements are available \emph{only} on \nancy. The BMC sampling frequency is 0.2~Hz (5~s period) on both clusters, while the PDU samples at 1~Hz (1~s period). 
As described in \cref{sec:background} and illustrated in \Cref{fig:illustration-energy-formula}, RAPL counters measure CPU-level energy consumption during kernel compilation. Since RAPL reports cumulative energy, the energy consumed over a build interval $[t_1, t_2]$ is obtained as $E_{\mathrm{RAPL}} = E_{\mathrm{RAPL}}(t_2) - E_{\mathrm{RAPL}}(t_1)$, which corresponds to the hatched area in \Cref{fig:illustration-energy-formula}. We read the RAPL package and DRAM counters through the Linux \texttt{perf\_event} interface (power PMU), opening one counter per socket and summing them. BMC and PDU energy ($E_{BMC}$ and $E_{PDU}$) are instead computed by integrating their respective power readings over the same $[t_{1},t_{2}]$ interval, as shown in \Cref{fig:illustration-energy-formula}. Additionally, we record compilation duration ($t_2-t_1$) and BMC-reported node temperatures (see \cref{app:temperature}).
The load induced by kernel compilation is incorporated into each node's baseline power consumption. Assuming a constant baseline~\cite{GuegainQR21} and identical configurations within each cluster, this contribution does not confound comparisons across tested kernel configurations.

\subsection{Evaluation Phase 1: Stability and Sensor Correlation (RQ1--RQ2, partial RQ3)}
\label{subsec:eval1}

The first evaluation phase is designed to answer RQ1 (framework calibration), RQ2 (sensor correlation), and to establish an initial assessment of the relationship between energy and build characteristics (RQ3). It uses all ten nodes across both clusters and covers a representative set of 15 Kconfig configurations:


\begin{itemize}
  \item \defconfig: the default configuration, used as the baseline.
  \item \texttt{tinyconfig}: a minimal configuration enabling only what is strictly necessary to produce a kernel.
  \item \texttt{allnoconfig}: disables every option that can be disabled.
  \item \allmodconfig: enables every loadable module. 
  \item \allyesconfig: enables every possible option. 
  \item \texttt{randconfig\_1--10}: ten random configurations generated from a fixed seed (\texttt{0xDEADBEEF}) for reproducibility.
\end{itemize}


This yields a set $C$ of $|C| = 15$ configurations spanning a wide range of compilation durations and kernel sizes, which correlate directly with the number of enabled options~\cite{BorgesPKA25} and thus provide a basis for studying energy variability. Each configuration is built 25 times on each of the 10 nodes, for a total of 3,750 builds. 
As shown in \Cref{fig:confvert-methodo}, prior to the initial build on each node, a fresh Debian OS is deployed and the required software dependencies (Docker and Sysbench) are installed. Each build then starts from a clean state: the kernel source tree is freshly copied before every build and deleted afterwards, ensuring no artifacts or cached objects persist across configurations or repetitions. Before the first build on each node, we execute a 2-minute Sysbench CPU benchmark using all available hardware threads as a warm-up phase, following recommendations from the literature \cite{KhanHNNO18}, bringing the processor to a thermally steady state and improving the stability of subsequent RAPL measurements.


\subsection{Evaluation Phase 2: Generalization Across Configurations (RQ3)}
\label{subsec:eval2}

The second evaluation phase targets the generalization of the energy--build-characteristics relationship established in Phase 1 (RQ3). Rather than repeating builds, it broadens the configuration space by sampling 200 distinct random Kconfig configurations, thereby covering a larger and more diverse region of the kernel configuration landscape. This phase runs on three \nancy nodes (\texttt{gros-100}, \texttt{gros-102}, and \texttt{gros-104}), with a single build per configuration per node. Random configurations were generated using \texttt{randconfig}; however, unlike in Phase 1 where all ten random configurations compiled successfully, a non-negligible fraction of configurations failed to build in this broader sample. Failed configurations were excluded and replaced with additional random draws until 200 passing configurations were obtained; investigating the causes of these build failures is left as future work. The same node setup procedure and clean-state policy from Phase 1 apply here (\Cref{fig:confvert-methodo}). This phase contributes 600 builds in total (200 configurations $\times$ 3 machines).
\section{Results}
\label{sec:results}

We present our results in three parts, following the structure of our research questions. RQ1 (\cref{sec:rq1.1,sec:rq1.2}) assesses the calibration and reproducibility of our framework. RQ2 (\cref{sec:rq2}) studies the agreement between RAPL, BMC, and PDU. RQ3 (\cref{sec:rq3}) evaluates build time and kernel size as energy proxies.

\subsection{RQ1 — Framework Calibration and Experimental Reproducibility}

We assess machine interchangeability (RQ1.1) and the minimum number of runs for reliable measurements (RQ1.2). In all comparisons, we exclude the two shortest builds, \texttt{tinyconfig} and \texttt{allnoconfig} (25--44\,s), which last only a few BMC sampling periods and cannot be measured reliably by that sensor. All comparisons therefore cover 13 configurations. While this section focuses on energy, build-time variability is discussed in \cref{app:build_time}.

\subsubsection{RQ1.1 — Cross-machine and cross-cluster interchangeability.}
\label{sec:rq1.1}

We ask whether machines within the same cluster give the same measurement — either in absolute value or up to a constant scaling factor — and whether results from one cluster carry over to the other, for all three energy metrics: RAPL (CPU+DRAM), BMC, and PDU (\nancy only).

\paragraph{\textbf{Equivalence test.}}
We consider two machines interchangeable when their energy for a given configuration differs by less than a margin $\delta$ (1, 2, or 5\%). A classical test of difference is not suited to this question: with 25 stable runs it detects even negligible gaps, and a non-significant result would not prove equivalence. We instead use the two one-sided test procedure (TOST)~\citep{schuirmann1987comparison,lakens2017equivalence}: a pair is equivalent at $\delta$ when the 90\% confidence interval of the difference lies within $\pm\delta$ ($\alpha = 5\%$). For each pair of machines, configuration, and metric, we compare their 25 runs on a log scale (so that differences are relative) using Welch's $t$-test~\citep{welch1947generalization,lakens2017equivalence}, which does not assume equal variances. We test two hypotheses:

\noindent $\bullet$ \emph{Absolute}: both machines give the same value.\\
\noindent $\bullet$ \emph{Up to a constant factor}: one machine may read systematically higher than the other (\eg by 8\% across all configurations). This offset affects absolute values but not rankings or ratios between configurations, and can be corrected once estimated. We remove it before the test by scaling one machine by the median of their difference over the \emph{other} configurations, so that a configuration never calibrates itself.

A pair is declared equivalent only if it is equivalent on all 13 configurations:  an intersection–union test~\citep{berger1996bioequivalence,casella2024statistical} that requires no multiple-testing correction when each individual test has level $\alpha$.

\begin{table}[t]
  \centering
  \small
  \caption{Pairs of machines (10 pairs from 5 machines per cluster) equivalent on all 13 configurations (\texttt{tinyconfig} and \texttt{allnoconfig} excluded), per energy metric (TOST, $\mathbf{\alpha = 5\%}$), in absolute value and up to a constant factor.}
  \label{tab:tost_pairs}
  \begin{tabular}{llrrrrrr}
    \toprule
    & & \multicolumn{3}{c}{Absolute} & \multicolumn{3}{c}{Constant factor} \\
    \cmidrule(lr){3-5}\cmidrule(lr){6-8}
    Cluster & Metric & $\pm1\%$ & $\pm2\%$ & $\pm5\%$ & $\pm1\%$ & $\pm2\%$ & $\pm5\%$ \\
    \midrule
    \rennes (C1) & RAPL & 0 & 2 & 10 & 0 & 10 & 10 \\
       & BMC          & 0 & 0 & 3  & 0 & 0  & 6  \\
    \midrule
    \nancy (C2) & RAPL & 0 & 3 & 6  & 0 & 7  & 10 \\
       & PDU          & 0 & 0 & 3  & 0 & 5  & 10 \\
       & BMC          & 0 & 0 & 2  & 0 & 0  & 3  \\
    \bottomrule
  \end{tabular}
\end{table}

\begin{figure*}[t]
  \centering
  \includegraphics[width=1\linewidth]{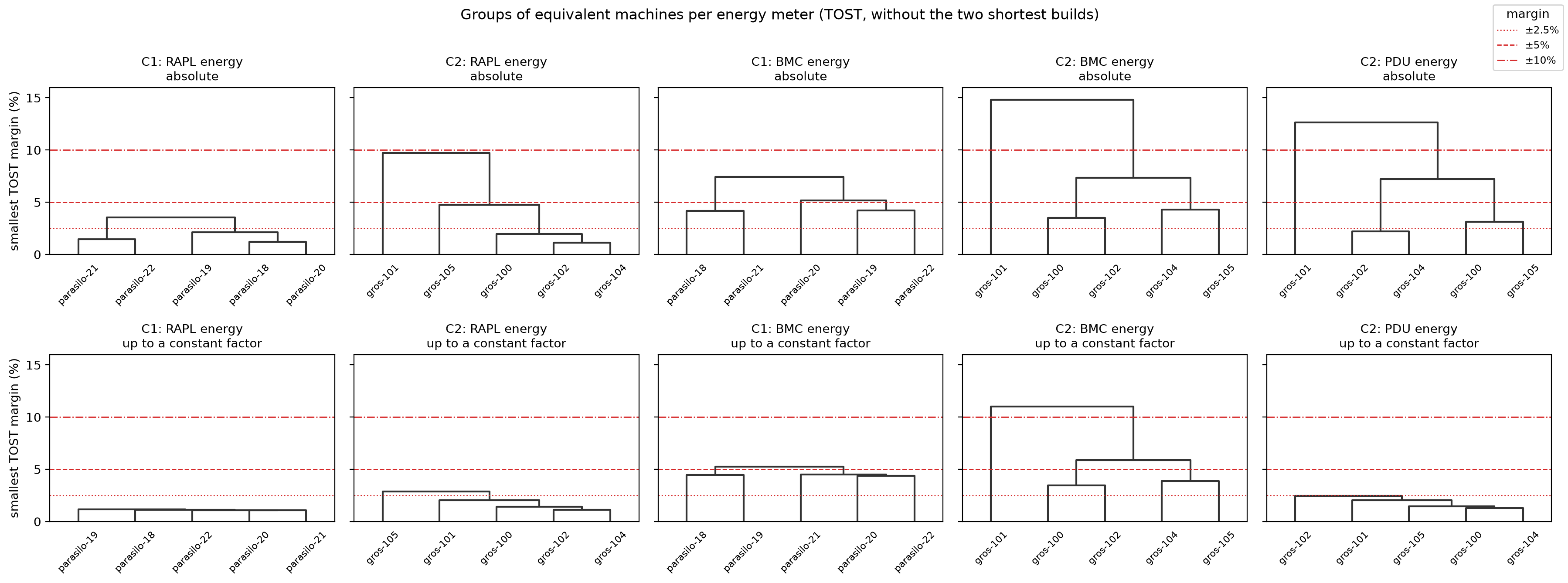}
  \caption{Groups of equivalent machines per energy metric and cluster (RAPL on \rennes (C2) and \nancy (C1), BMC on \rennes and \nancy, PDU on \nancy only), in absolute value (top) and up to a constant factor (bottom). The merge height is the smallest TOST margin at which every pair is equivalent on all 13 configurations (\texttt{tinyconfig} and \texttt{allnoconfig} excluded; complete linkage). All panels share the same $\mathbf{y}$-axis. Red lines mark $\mathbf{\pm 2.5}$, $\mathbf{5}$, and $\mathbf{10\%}$ margins.}
  \label{fig:tost_groups}
  \Description{Groups of equivalent machines per energy metric and cluster (RAPL, BMC and PDU), in absolute value (top) and up to a constant factor (bottom). The merge height is the smallest TOST margin at which every pair is equivalent on all 13 configurations (tinyconfig and allnoconfig excluded; complete linkage).}
\end{figure*}

\paragraph{\textbf{Within a cluster.}}
\Cref{tab:tost_pairs} reports equivalent pairs at each margin. No pair is equivalent at $\pm 1\%$ in absolute value; at $\pm 2\%$, at most 3 of 10 pairs pass, and on \nancy, even $\pm 5\%$ fails for 4 RAPL and 7 PDU pairs. Removing the constant factor makes all pairs equivalent at $\pm 5\%$ for RAPL on both clusters and for the PDU. On \rennes, all RAPL pairs pass at $\pm 2\%$, as do 7 RAPL and 5 PDU pairs on \nancy. These factors (up to $8.0\%$ for RAPL and $9.2\%$ for the PDU on \nancy) are non-negligible but configuration-independent and thus correctable. BMC remains less consistent: only 6 pairs on \rennes and 3 on \nancy pass at $\pm 5\%$ after correction.

\Cref{fig:tost_groups} groups interchangeable machines using the smallest margin at which two machines are equivalent across all configurations for a given metric, with complete linkage ensuring pairwise equivalence within each group. For RAPL, all \rennes machines merge at $3.5\%$ in absolute value and $1.2\%$ after correction. On \nancy, \texttt{gros-101} remains separate up to $9.7\%$ in absolute value, while the other four merge at $4.8\%$; after correction, all merge at $2.9\%$. For BMC, all \rennes machines merge at $7.4\%$ in absolute value and $5.3\%$ after correction. On \nancy, \texttt{gros-101} remains separate up to $14.8\%$ and $11.0\%$, respectively, while the other four merge at 7.3\% in absolute value and 5.9\% after correction. For the PDU (\nancy only), \texttt{gros-101} remains separate up to $12.6\%$ in absolute value, while the other machines merge at $7.2\%$; after correction, all merge at $2.4\%$, the tightest \nancy grouping. \texttt{gros-101} is thus an outlier on all three metrics in absolute value. Its RAPL and PDU offsets are correctable, whereas its BMC measurements remain distinct even after correction.





\paragraph{\textbf{Across clusters.}}

\begin{figure}[t]
  \centering
  \includegraphics[width=1\linewidth]{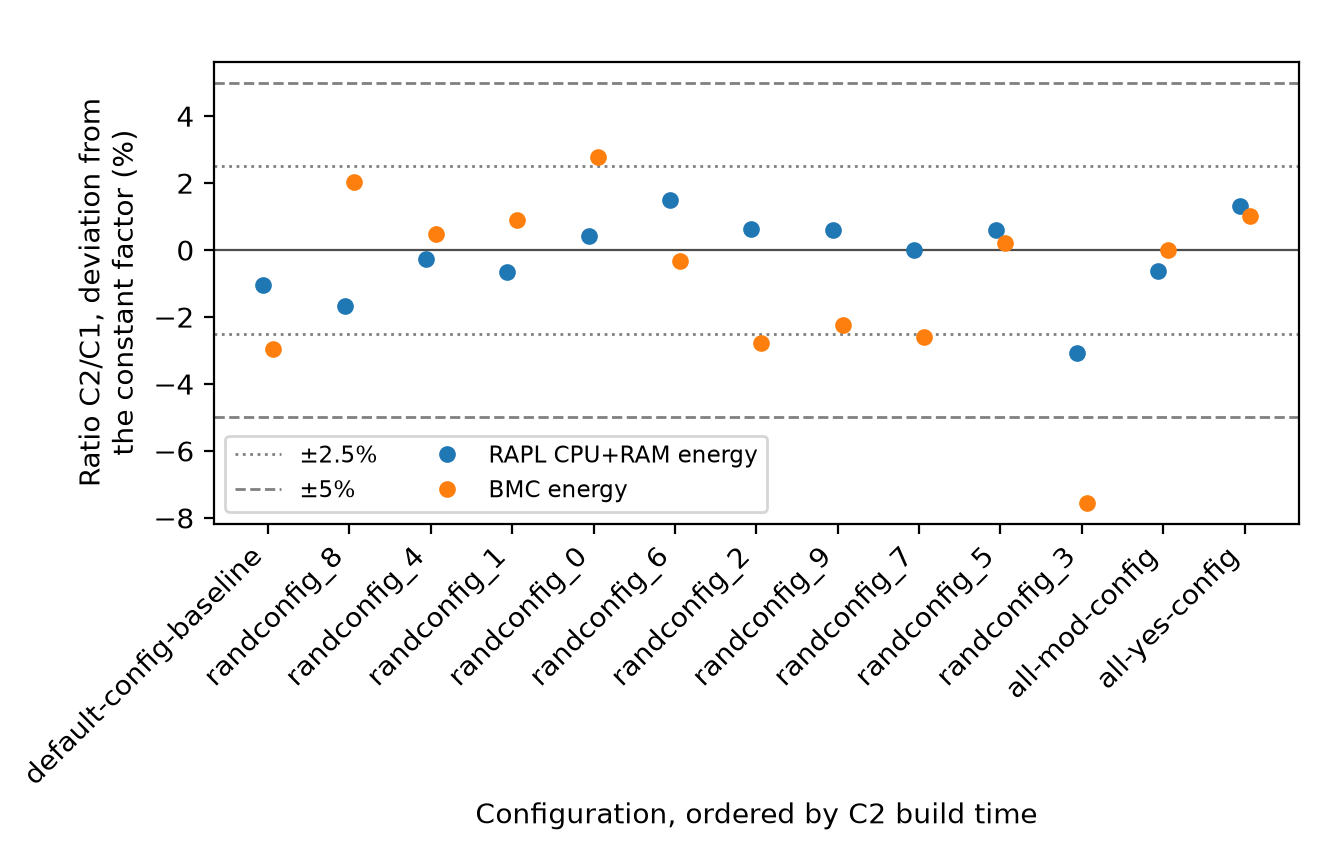}
  \caption{Deviation of the \nancy (C2)/\rennes (C1) ratio of each of the 13 configurations from the constant factor between clusters (median ratio), for RAPL (CPU+DRAM) and BMC energy. Configurations are ordered by build time on \nancy; grey lines mark the $\mathbf{\pm 2.5\%}$ and $\mathbf{\pm 5\%}$ margins.}
  \label{fig:cross_cluster}
  \Description{Deviation of the \nancy (C2)/\rennes (C1) ratio of each of the 13 configurations from the constant factor between clusters (median ratio), for RAPL (CPU+DRAM) and BMC energy.}
\end{figure}

The two clusters differ in hardware, so only relative results can transfer. We compare RAPL and the BMC, both available on \rennes and \nancy. \Cref{fig:cross_cluster} shows the deviation of the \nancy/\rennes ratio of each configuration from the constant factor between clusters (median ratio over configurations). For RAPL, every configuration is within $3.1\%$ of this factor; for the BMC, within $7.5\%$. Rankings are preserved as well: Kendall's $\tau_b$ between the configuration means of \rennes and \nancy is $1.00$ for RAPL (the 13 configurations are ranked identically) and $0.97$ for the BMC (a single discordant pair out of 78). A TOST across clusters (5 machine means per side) confirms the factor at $\pm 5\%$ for 11/13 configurations with RAPL and 8/13 with the BMC; at $\pm 2.5\%$, power is insufficient to confirm any configuration, consistent with the observed deviations being smaller than $\pm 5\%$ rather than larger.



\subsubsection{RQ1.2 — Minimum number of runs.} \label{sec:rq1.2}

We quantify measurement reliability through the \emph{precision of the mean}: the half-width of the 95\% confidence interval of the mean of $n$ runs, relative to the mean. We study each configuration on each machine separately, retain the worst machine, and summarize over configurations with the median (typical case) and the worst (guarantee for all configurations).

\paragraph{How many runs are sufficient?}
To obtain a reusable rule for new configurations, we apply the classical sample-size formula for a mean~\citep{jain1990art}: the minimum $n$ such that
\begin{equation}
  t_{0.975,\,n-1} \cdot \frac{\mathit{CV}}{\sqrt{n}} \le e
  \label{eq:sample_size}
\end{equation}
where $e$ is the target half-width (5 or 10\% of the mean) and $\mathit{CV}$ is the run-to-run coefficient of variation estimated from the 25 runs. We use the Student-$t$ quantile rather than the normal one, as recommended by \citet{georges2007statistically}: in a future campaign, the interval of each configuration will be estimated from its own $n$ runs, so using $t$ is intentionally conservative ($t_{0.975,1} = 12.7$ and $t_{0.975,2} = 4.3$, against $1.96$ for the normal quantile). Because $t$ depends on $n$, we solve \Cref{eq:sample_size} by checking $n = 2, 3, \dots$ until the inequality holds. 
\Cref{tab:runs_needed} gives the results.


\begin{table}[t]
  \centering
  \small
  \caption{Minimum number of runs per configuration (worst machine) for a 95\% CI within $\mathbf{\pm 5\%}$ and $\mathbf{\pm 10\%}$ of the mean (\Cref{eq:sample_size}). Median: median configuration; Worst: worst configuration. Last column: share of the 13 configurations reaching $\mathbf{\pm 5\%}$ with at most 5 runs.}
  \label{tab:runs_needed}
  \begin{tabular}{llrrrrr}
    \toprule
    & & \multicolumn{2}{c}{$\pm 5\%$} & \multicolumn{2}{c}{$\pm 10\%$} & \\
    \cmidrule(lr){3-4}\cmidrule(lr){5-6}
    Cluster & Metric & Median & Worst & Median & Worst & $\le 5$ runs ($\pm 5\%$) \\
    \midrule
    \rennes & RAPL & 2 & 4   & 2 & 3  & 100\% \\
       & BMC          & 3 & 17  & 3 & 7  & 92\% \\
    \midrule
    \nancy & RAPL & 2 & 4   & 2 & 3  & 100\% \\
       & BMC          & 5 & 8 & 3 & 4 & 54\% \\
       & PDU          & 2 & 4   & 2 & 3  & 100\% \\
    \bottomrule
  \end{tabular}
\end{table}

For RAPL and the PDU, 2 runs yield a $\pm 5\%$ interval for the typical configuration; 4 runs suffice for every configuration on both clusters. Reducing the protocol from 25 to 4 runs would thus cut Phase~1 from 3,750 to 600 builds (84\% fewer) while keeping a $\pm 5\%$ interval for RAPL and the PDU. The worst case is always \defconfig, whose run-to-run CV is 2.0--2.6\% against a median of 0.3--0.4\%, due to occasional outlier runs. 

The BMC requires more: 3 (\rennes) to 5 (\nancy) runs for the typical configuration. On \rennes, the worst case is again \defconfig: a single machine (parasilo-19, CV $9.6\%$) requires 17 runs, while every other configuration reaches $\pm 5\%$ within 4 runs. On \nancy, no single configuration stands out: the run-to-run CV of the BMC is higher overall (median $3.9\%$), and 6 of the 13 configurations need 6 to 8 runs at $\pm 5\%$. The worst ones are short builds (2--3\,min), which cover fewer 5\,s BMC periods. At $\pm 10\%$, 4 runs suffice for every configuration on \nancy, and 7 on \rennes.

\vspace{-0.6\baselineskip}
\paragraph{\textbf{Runs versus machines.}}
More runs on one machine reduce run-to-run noise but not the between-machine offset identified in RQ1.1. Given a fixed budget of $m \times r$ measurements (for $r$ runs on $m$ machines), the variance of the cluster mean is

\begin{equation}
  \operatorname{Var}(\bar{E}) = \frac{\sigma^2_\text{machine}}{m} + \frac{\sigma^2_\text{run}}{m\,r}
  \label{eq:var_mean}
\end{equation}
More runs only shrink the second term. For RAPL and the PDU, the machine-to-machine CV is 1.1--3.8\% against 0.2--0.3\% between runs, so 96--99.7\% of the variance comes from the machine (ICC). For example, with CVs of 3.8\% and 0.3\%, 25 runs on one machine yield a relative standard error of 3.8\%, against 1.7\% for a single run on each of 5 machines. Any budget saved on runs is thus better spent on additional machines. The optimal runs-per-machine~\citep{kalibera2013rigorous} is given by
\begin{equation}
  R_\text{opt} = \left\lceil \sqrt{c \cdot \sigma^2_\text{run} / \sigma^2_\text{machine}} \right\rceil
\end{equation}
where $c$ is the cost of adding a machine relative to adding a run. $R_\text{opt} = 1$ as long as $c \le \sigma^2_\text{machine}/\sigma^2_\text{run}$, \ie for RAPL and the PDU, as long as a machine costs less than $\approx$25 runs on \rennes or $\approx$150 on \nancy. For the BMC, whose run-to-run noise matches its machine-to-machine differences, $R_\text{opt}$ is 3--4 for a cost ratio of 5. This applies to absolute energies; for rankings or ratios between configurations on the same machine, the constant factor cancels and \Cref{tab:runs_needed} applies directly.

\SumupBoxViolet[Answer to RQ1]{
Machines with identical hardware are not interchangeable in absolute energy, but they are up to a stable per-machine factor (up to $8.0\%$ for RAPL and $9.2\%$ for the PDU): after correction, \emph{all} pairs are equivalent at $\pm 5\%$ for RAPL and the PDU. Relative conclusions also transfer across clusters: RAPL preserves configuration rankings and its \nancy/\rennes ratio remains within $3.1\%$ of a constant factor. RAPL and the PDU require few runs: 2 typically suffice for a 95\% CI within $\pm 5\%$, and 4 cover all tested configurations (84\% fewer builds than with 25 runs); the saved budget is better spent on additional machines, as 96--99.7\% of the variance stems from machine differences. The BMC is less reliable on every aspect: margins of $\pm 5.3\%$ (\rennes) and $\pm 11.0\%$ (\nancy) even after correction, deviations up to $7.5\%$ across clusters, and up to 17 (\rennes) and 8 (\nancy) runs.}

\subsection{RQ2 — Cross-Sensor Agreement Between RAPL, BMC, and PDU Measurements}
\label{sec:rq2}

\begin{figure*}
    \centering
    \includegraphics[width=1\linewidth]{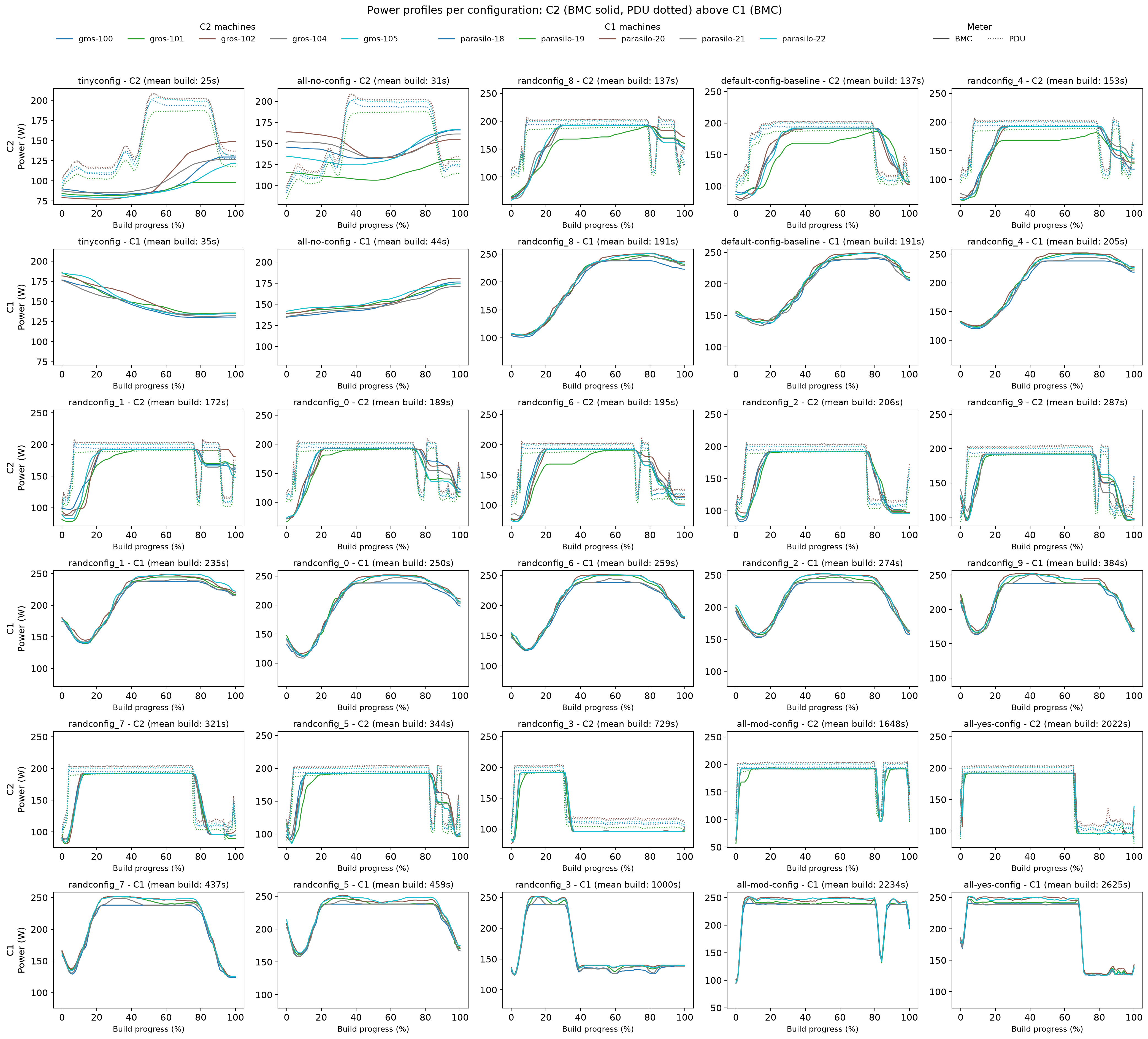}
    \caption{Mean power profiles per machine (one colour per machine) and per configuration, ordered by mean build time. Odd rows: \nancy (C2), BMC (solid) and PDU (dotted). Even rows: \rennes (C1), BMC only (no PDU available). Each curve averages the 25 runs of one machine after rescaling time to the percentage of build progress; each panel title gives that cluster's mean build duration.}
    \label{fig:bmc_vs_pdu_profiles}
    \Description{Mean power profiles per machine (one colour per machine) and per configuration, ordered by mean build time. Each curve averages the 25 runs of one machine after rescaling time to the percentage of build progress; each panel title gives that cluster's mean build duration.}
\end{figure*}


RQ1 showed that relative energy differences are robust across machines and clusters. RQ2 asks a different question: do sensors that measure different scopes and at different rates nonetheless agree? RAPL captures CPU and DRAM energy; the BMC (0.2\,Hz) and PDU (1\,Hz) integrate whole-node power over the build window. We compare all three on \nancy (the only cluster with a PDU) and RAPL against the BMC on both clusters. 

\paragraph{\textbf{BMC vs.\ PDU}}
\Cref{fig:bmc_vs_pdu_profiles} shows that for a given configuration, all machines in a cluster follow similar power profiles on both sensors. The sequence of build phases is shared across clusters: \texttt{randconfig\_3} alternates a short high-power phase with a long low-power one on both clusters, and \allmodconfig shows the same brief power drop at $\approx$80\% of the build. Only absolute power levels differ between clusters.

On \nancy, BMC and PDU profiles share the same shape for builds longer than one minute, but differ in two ways:\\
\noindent $\bullet$ \textbf{Level.} The PDU reads consistently above the BMC.\\
\noindent $\bullet$ \textbf{Timing.} The BMC lags behind the PDU on sharp transitions, such as the ramp-up at the start of the build.


A Bland--Altman analysis~\cite{bland1986statistical} on $100 \cdot (\mathrm{BMC} - \mathrm{PDU}) / \mathrm{mean}$ quantifies this for builds longer than one minute (13 configurations, 1,625 builds). The BMC energy is on average $7.0\%$ lower than the PDU energy, with 95\% agreement limits of $[-16.0\%, +2.0\%]$. The bias is consistently stable across configurations ($-4.2\%$ for \allmodconfig to $-10.8\%$ for \texttt{randconfig\_8}) and narrows with build duration (from $\approx$20\,W at 140\,s to $\approx$9\,W above 1,000\,s), suggesting a BMC edge effect rather than a purely instrumental offset.

Agreement collapses for the two shortest builds: biases of $-18.3\%$ (\texttt{allnoconfig}, 31\,s) and $-47.7\%$ (\texttt{tinyconfig}, 25\,s), with limits of agreement wider than $\pm 50$ points. As shown in the first two subplots of \Cref{fig:bmc_vs_pdu_profiles}, the PDU — sampling five times more frequently — captures the power ramp at build start, while the BMC lags and smooths it, causing a substantial fraction of energy to fall outside the integration window. BMC measurements for builds shorter than $\approx$1 minute should therefore not be trusted.

\begin{figure*}
    \centering
    \includegraphics[width=0.97\linewidth]{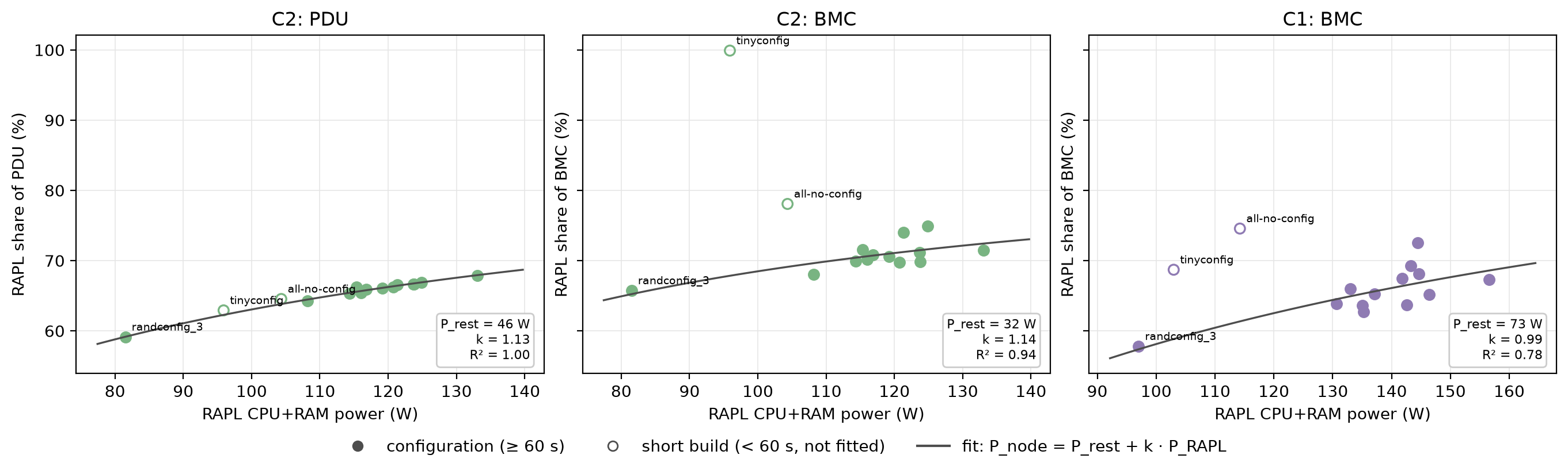}
    \caption{RAPL (CPU+DRAM) share of the whole-node power measured by the BMC (\rennes (C1), \nancy (C2)) and the PDU (\nancy (C2)), one point per configuration (median over 125 builds). The curve is the share implied by the fit $\mathbf{P_\text{\textbf{node}} = P_\text{\textbf{rest}} + k \cdot P_\text{\textbf{RAPL}}}$ (values in each panel). Hollow points are builds shorter than 60\,s, left out of the fit.}
    \label{fig:rapl_vs_bmc_pdu}
    \Description{RAPL (CPU+DRAM) share of the whole-node power measured by the BMC and the PDU, one point per configuration (median over 125 builds). The curve is the share implied by a proportional fit (values in each panel). Hollow points are builds shorter than 60\,s, left out of the fit.}
\end{figure*}

\paragraph{\textbf{RAPL vs.\ system-level metrics.}}
We model whole-node power as a fixed baseline plus a term proportional to RAPL:
\begin{equation}
  P_\text{system} = P_\text{rest} + k \cdot P_\text{RAPL}
  \label{eq:system_rapl}
\end{equation}
where $P_\text{rest}$ is the power drawn outside CPU and DRAM (fans, disks, network, mainboard), and $k$ is the system watts per RAPL watt. We fit \Cref{eq:system_rapl} by ordinary least squares on the median average power of each configuration, leaving out the two short builds (\Cref{fig:rapl_vs_bmc_pdu}).

With the PDU the fit is almost exact ($R^2 > 0.99$, residual standard deviation below 1\,W): $P_\text{rest} = 46$\,W (95\% CI $[42, 50]$) and $k = 1.13$ $[1.09, 1.16]$. In other words, each RAPL watt costs $\approx$1.13\,W at the wall, on top of a fixed 46\,W. Even the two short builds, excluded from the fit, lie on the regression line. The BMC yields a consistent but noisier picture: on \nancy, $P_\text{rest} = 32$\,W $[9, 55]$ and $k = 1.14$ $[0.94, 1.34]$ ($R^2 = 0.94$); on \rennes, $P_\text{rest} = 73$\,W $[24, 121]$ and $k = 0.99$ $[0.64, 1.35]$ ($R^2 = 0.78$). Since the PDU explains virtually all between-configuration differences, the lower BMC $R^2$ most plausibly reflects measurement noise. On \nancy the BMC and PDU share the same slope ($k \approx 1.14$) and differ only in intercept ($\approx$14\,W), independently supporting an offset reading of the BMC--PDU gap. A non-zero $P_\text{rest}$ implies that the RAPL share of system energy is not constant: it decreases when builds draw less CPU power, as the fixed baseline then weighs more. Concretely, RAPL accounts for 58--68\% of PDU energy across configurations (CV 3.5\%), 66--74\% of BMC energy on \nancy, and 58--73\% on \rennes. The lowest share always belongs to \texttt{randconfig\_3}, the configuration with the lowest average RAPL power. On the BMC, the two short builds appear to reach up to 100\% (\texttt{tinyconfig}): this is the sampling artefact described above, not a property of the build, as confirmed by their position on the PDU regression line.


\SumupBoxViolet[Answer to RQ2]{BMC and PDU track the same power dynamics but differ systematically: the BMC reads on average 7.0\% lower than the PDU (limits of agreement $[-16.0\%, +2.0\%]$), with a bias combining a constant offset and an edge effect that weighs less for longer builds. For builds shorter than $\approx$1 minute, the BMC misses a substantial fraction of energy and is not trustworthy. RAPL captures 58--68\% of whole-node PDU energy via a near-perfect linear relationship 
($P_\text{PDU} = P_\text{rest} + k \cdot P_\text{RAPL}$; $R^2 > 0.99$, $k \approx 1.13$, $P_\text{rest} \approx 46$\,W), 
confirming that RAPL reliably tracks \emph{relative} differences between configurations even though it systematically underestimates absolute node energy.}

\subsection{RQ3 — Build Duration and Kernel Size as Proxies for Energy Consumption} \label{sec:rq3}

We assess whether build time or kernel size (\ie the size of the compiled \texttt{vmlinux} binary) consistently ranks and estimates configuration energy, first on the 15 configurations of Phase~1 (both clusters), then on the 200 configurations of Phase~2 (\nancy only) to test generalizability.

 

\paragraph{\textbf{Rank correlation.}}
We measure how consistently each proxy orders configurations by energy using Kendall's $\tau_b$. Since the 125 builds of a configuration (5 machines $\times$ 25 runs) are repetitions, we compute $\tau_b$ on configuration medians (15 points, 105 pairs per cluster). A pair is concordant when the larger proxy corresponds to higher energy, so $\tau_b = (C-D)/105$, with $C$ and $D$ the numbers of concordant and discordant pairs. \Cref{tab:rq3-kendall} reports exact $p$-values adjusted with Holm's procedure over the 10 tests (for Phase~1), and 95\% bootstrap CIs obtained by resampling configurations.

All correlations remain significant after Holm correction ($p_\text{Holm} < 10^{-4}$), but their strength differs markedly. Build time perfectly ranks configurations for BMC (both clusters) and PDU (\nancy), with no discordant pairs among the 105 ($\tau_b = 1.00$). For RAPL, only one pair is discordant on \nancy ($\tau_b = 0.981$): \allmodconfig and \allyesconfig, as the latter builds longer despite similar average RAPL power. On \rennes, two pairs are discordant ($\tau_b = 0.962$), including \texttt{randconfig\_0} and \texttt{randconfig\_6}, whose build times differ by only 9\,s. The \allmodconfig/\allyesconfig swap is consistent with the lower average CPU power of \allyesconfig, which builds longer while drawing about the same RAPL energy. 
Kernel size is a much weaker proxy: 12--14 pairs are discordant ($\tau_b =  0.733$--$0.771$, lower CI bound $\approx$0.5), spread across several random configurations that produce large kernels in comparatively short builds (\eg \texttt{randconfig\_7} and \texttt{randconfig\_9}). Since BMC and PDU energy follow the build-time order exactly, their $\tau_b$ with kernel size equals that of build time vs.\ kernel size ($\tau_b = 0.771$): kernel size carries no ranking information beyond what build time already provides. 

On the 200 configurations of Phase~2, build time remains a strong proxy ($\tau_b = 0.846$--$0.888$), but far from perfect (1,113--1,536 discordant pairs out of 19,900). Kernel size drops sharply ($\tau_b = 0.286$--$0.308$, $\approx$7,000 discordant pairs), confirming it is a weak proxy across a diverse configuration space.

\begin{table}[t]
  \setlength{\tabcolsep}{2pt}
  \centering
  \small
  \caption{Rank agreement between proxies and energy metrics on Phase~1 (top, $\mathbf{n=15}$ configurations, 105 pairs per cluster) and Phase~2 (bottom, $\mathbf{n=200}$ configurations, 19,900 pairs, \nancy only): Kendall's $\mathbf{\tau_b}$ with 95\% bootstrap CI, discordant pairs, and Holm-corrected $\mathbf{p}$-value.}
  \label{tab:rq3-kendall}
  \begin{tabular}{lllcccc}
    \toprule
    Proxy & Measured & Cluster & $\tau_b$ & 95\% CI & Disc.\ pairs & $p_{\text{Holm}}$ \\
    \midrule
    \multicolumn{7}{c}{\textbf{\emph{Phase 1 — 15 configurations}}} \\
    \multirow{5}*{Kernel size}
      & \multirow{2}*{BMC}  & \nancy  & 0.771 & [0.54, 0.96] & 12/105 & $7.0\times10^{-5}$ \\
      &&                       \rennes & 0.771 & [0.54, 0.96] & 12/105 & $7.0\times10^{-5}$ \\
      & PDU                 & \nancy  & 0.771 & [0.54, 0.96] & 12/105 & $7.0\times10^{-5}$ \\
      & \multirow{2}*{RAPL} & \nancy  & 0.752 & [0.52, 0.93] & 13/105 & $7.0\times10^{-5}$ \\
      &&                       \rennes & 0.733 & [0.49, 0.92] & 14/105 & $7.0\times10^{-5}$ \\
    \hdashline
    \multirow{5}*{Build time}
      & \multirow{2}*{BMC}  & \nancy  & 1.000 & [1.00, 1.00]$^\dagger$ & 0/105 & $1.8\times10^{-11}$ \\
      &&                       \rennes & 1.000 & [1.00, 1.00]$^\dagger$ & 0/105 & $1.8\times10^{-11}$ \\
      & PDU                 & \nancy  & 1.000 & [1.00, 1.00]$^\dagger$ & 0/105 & $1.8\times10^{-11}$ \\
      & \multirow{2}*{RAPL} & \nancy  & 0.981 & [0.75, 1.00] & 1/105  & $2.1\times10^{-10}$ \\
      &&                       \rennes & 0.962 & [0.83, 1.00] & 2/105  & $1.5\times10^{-9}$  \\
    \midrule
    \multicolumn{7}{c}{\textbf{\emph{Phase 2 — 200 configurations (\nancy only)}}} \\
    \multirow{3}*{Kernel size}
      & BMC  & \nancy & 0.299 & [0.21, 0.38] & 6{,}972/19{,}900 & $6.2\times10^{-10}$ \\
      & PDU  & \nancy & 0.308 & [0.22, 0.38] & 6{,}885/19{,}900 & $2.8\times10^{-10}$ \\
      & RAPL & \nancy & 0.286 & [0.20, 0.36] & 7{,}106/19{,}900 & $1.9\times10^{-9}$  \\
    \hdashline
    \multirow{3}*{Build time}
      & BMC  & \nancy & 0.866 & [0.84, 0.89] & 1{,}338/19{,}900 & $2.4\times10^{-122}$ \\
      & PDU  & \nancy & 0.888 & [0.86, 0.91] & 1{,}113/19{,}900 & $2.8\times10^{-136}$ \\
      & RAPL & \nancy & 0.846 & [0.81, 0.87] & 1{,}536/19{,}900 & $6.1\times10^{-112}$ \\
    \bottomrule
  \end{tabular}
  
  \footnotesize $^\dagger$ No discordant pair in any bootstrap resample: CI degenerates to $[1, 1]$.
\end{table}

\paragraph{\textbf{Estimation accuracy.}}
Since kernel size consistently underperforms build time as a ranking proxy, we focus the estimation analysis on build time alone. Ranking and estimation are distinct: even a high $\tau_b$ does not imply accurate energy prediction. To quantify the latter, we use leave-one-configuration-out cross-validation on the 200 configurations of Phase~2: for each configuration in turn, we fit a model on the remaining 199 medians and predict its energy from build time alone, mimicking the intended use case of estimating the energy of an unmeasured configuration. We compare three models of increasing flexibility: proportional ($E = P\,t$, a single average power $P$), linear ($E = a + b\,t$), and log-log ($\log E = a + b \cdot \log t$). Results for \nancy are shown in \Cref{tab:rq3-accuracy}.

The proportional model achieves the lowest median error across all metrics (3.5\% for the PDU, 4.5\% for the BMC, 4.7\% for RAPL); the log-log model is close behind (3.9--5.5\%) and the linear model is always higher (4.5--6.3\%). These typical errors exceed the spread between machines of the cluster for every metric: about 1.5 times the machine CV for the BMC and the PDU (2.8\% and 2.5\%), and about 14 times for RAPL (0.3\%) Estimating energy from build time is thus less accurate than measuring it on another machine of the same cluster. The worst cases are large (up to 37\% for the BMC, 35\% for the PDU and 52\% for RAPL) and come from two configurations, \texttt{randconfig\_62} and \texttt{randconfig\_123}, the two longest builds of Phase~2 (11--12\,min). Both are overestimated: they draw full power for only the first 3\,min, then stay close to idle power for the remaining 8\,min, so their duration overstates their energy. Very short builds suffer from the same bias in the opposite way: they spend a larger share of their run below full power, so a single average power overestimates their energy. Phase~2 contains no build under 73\,s, but in Phase~1 the proportional model overestimates \texttt{tinyconfig} (25--35\,s) by up to 72\% for the BMC and 35\% for RAPL. The BMC is hit hardest because it takes about 20\,s to reflect a change in power. Build time is therefore a reliable energy estimate for builds that keep the machine busy for most of their run, but not for very short builds or for builds with a long low-power tail.

\begin{table}[t]
  \setlength{\tabcolsep}{4pt}
  \centering
  \small
  \caption{Leave-one-out energy estimation error (\%) from build time, over 200 configurations (\nancy, Phase~2): proportional ($\mathbf{E = P\,t}$, with $\mathbf{P}$ the median
  average power), linear ($\mathbf{E = a + b\,t}$), and log-log ($\mathbf{\log E = a + b \cdot \log t}$) models fitted on configuration medians. 
  \emph{Machines}: CV of a single build on another machine of the cluster (machine and run-to-run variance combined). Unlike \cref{sec:rq1.1}, Phase~2 considers only three close machines of \nancy (\texttt{gros-100}, \texttt{gros-102}, \texttt{gros-104}) and one run per machine.}
  \label{tab:rq3-accuracy}
  \begin{tabular}{lccccccc}
    \toprule
    & \multicolumn{2}{c}{Proportional} & \multicolumn{2}{c}{Linear} & \multicolumn{2}{c}{Log-log} & \\
    \cmidrule(lr){2-3}\cmidrule(lr){4-5}\cmidrule(lr){6-7}
    Metric & Median & Max & Median & Max & Median & Max & Machines \\
    \midrule
    BMC & \cellcolor{pastelgreen}{4.5} & 36.8 & 5.4 & 38.0 & 4.6 & 33.0 &  2.84 \\
    PDU & \cellcolor{pastelgreen}{3.5} & 35.0 & 4.5 & 26.9 & 3.9 & 29.3 &  2.49 \\
    RAPL & \cellcolor{pastelgreen}{4.7} & 51.6 & 6.3 & 38.0 & 5.5 & 42.0 &  0.34 \\
    \bottomrule
  \end{tabular}
\end{table}

\SumupBoxViolet[Answer to RQ3]{Build time is a reliable proxy for \emph{ranking} configurations by energy: on 15 configurations (Phase~1) it orders all 105 pairs correctly for whole-node metrics ($\tau_b = 1.00$) and at least 103/105 for RAPL. This strong ranking holds broadly on 200 configurations (Phase~2, $\tau_b = 0.85$--$0.89$), though with more exceptions as the configuration space diversifies. Build time is only a rough \emph{estimator} of absolute energy: typical leave-one-out error is 3.5--4.7\%, exceeding the between-machine spread for BMC and PDU, and reaching up to 72\% for atypical power profiles or very short builds. Kernel size is a weak proxy throughout: it misorders 12--14 pairs on 15 configurations and $\approx$7,000 on 200, adding no information beyond build time.}

\section{Threats to Validity}
\label{sec:threats}

This study quantifies the energy consumption of Linux kernel builds across configurations and heterogeneous hardware and software environments. We ran our experiments on Grid'5000, which provides bare-metal node access, systematic OS reloading before each execution, and tooling that supports reproducibility. Despite these controls, several threats remain.



\paragraph{\textbf{Internal Validity.}} The hardware platform we used is not cutting-edge: the \nancy nodes (Dell PowerEdge R640) were deployed in September 2019 and the \rennes nodes (Dell PowerEdge R630) in October 2015.  Results may therefore be affected by server ageing, including silicon ageing phenomena that increase CPU power consumption over time~\cite{ChanCK14}.
CPU energy was measured with RAPL under real-world conditions (C-states and Turbo Boost enabled), so we cannot exclude that these features, or background processes, introduced noise in the measurements.

\paragraph{\textbf{External Validity.}} Our experiments cover only two server models equipped with two generations of Intel Xeon processors (two Xeon E5-2630~v3 and one Xeon Gold 5220). The results may therefore not generalize to other server models, whose power supplies, cooling, and on-board sensors differ or to AMD processors. Similarly, we consider a single kernel version (v7.0), a single target architecture (x86\_64), and a single compiler (GCC~12); other compilers such as Clang, or cross-compilation, could alter the energy profile of the builds. These limitations are inherent to deep software variability~\cite{LesoilABJ21}: each additional layer multiplies the number of settings to measure, and we leave the exploration of these layers to future work. The machines we used may eventually be retired from Grid'5000, which would prevent the reproduction of our experiments on the exact same hardware.

Finally, with tens of thousands of options, the Linux kernel yields an astronomical number of configurations. Phase~1 samples only 15 (five canonical presets and ten random draws), capturing representative extremes and typical cases but leaving most of the space unexplored. Phase~2 expands coverage to 200 random configurations, providing a broader sample that nevertheless remains negligible relative to the full space.

\paragraph{\textbf{Construct Validity.}} RAPL's package and DRAM domains do not account for storage, networking, or power-supply losses. Given the substantial I/O involved in a kernel build (\eg, copying the source tree and writing object files), RAPL likely underestimates actual energy consumption. This is one motivation for comparing it with BMC and PDU measurements, which capture the whole machine.
However, these external devices also have limitations: BMCs sample every 5\,s and PDUs every 1\,s, and build timestamps ($t_1$, $t_2$) are taken from a different clock than Kwollect samples. The resulting boundary errors are negligible for long builds but can be substantial for short ones such as \texttt{tinyconfig} ($\approx$30\,s).
Finally, PDU measurements are only available on \nancy, so BMC--PDU agreement is assessed on a single cluster and may not generalize to other hardware or BMC firmware. 
\section{Conclusion}
\label{sec:conclusion}

We presented a framework to measure the energy of building Linux kernel configurations on bare-metal nodes with RAPL, BMC and PDU, and evaluated it on 4,350 builds across two clusters.
Our goal is to calibrate energy measurements in order to then perform large-scale exploration of configuration build spaces.
Several notable findings emerge from this analysis.
For example, machines with identical hardware are not interchangeable in absolute energy, but they differ by stable factors, so rankings and ratios between configurations transfer across machines and clusters.
For RAPL and the PDU, two to four runs per configuration suffice, and the remaining budget is better spent on additional machines than on additional runs.
Our experiments also suggest some trends and hypotheses to further investigate.
In particular build time is a reliable proxy to rank configurations by energy, yet only a rough estimator of its value, while kernel size is a poor proxy.
Thanks to these results, we can now envision scaling to thousands of configurations and to investigate build-energy prediction and tuning, energy-aware continuous integration of configurations, and the identification of the configuration options that most affect build energy.

\section*{Use of Generative AI}
\label{app:genai}

The authors used a generative AI assistant (Claude, Anthropic) to help write some scripts (\eg to analyze the data), to review the manuscript for consistency, and to rephrase parts of the text. The authors checked the scripts and their outputs, reviewed all the content, and take full responsibility for the paper.

\newpage

\newpage
\appendix
\section{Ambient temperature during the builds}
\label{app:temperature}

\begin{figure*}[t]
  \centering
  \includegraphics[width=\textwidth]{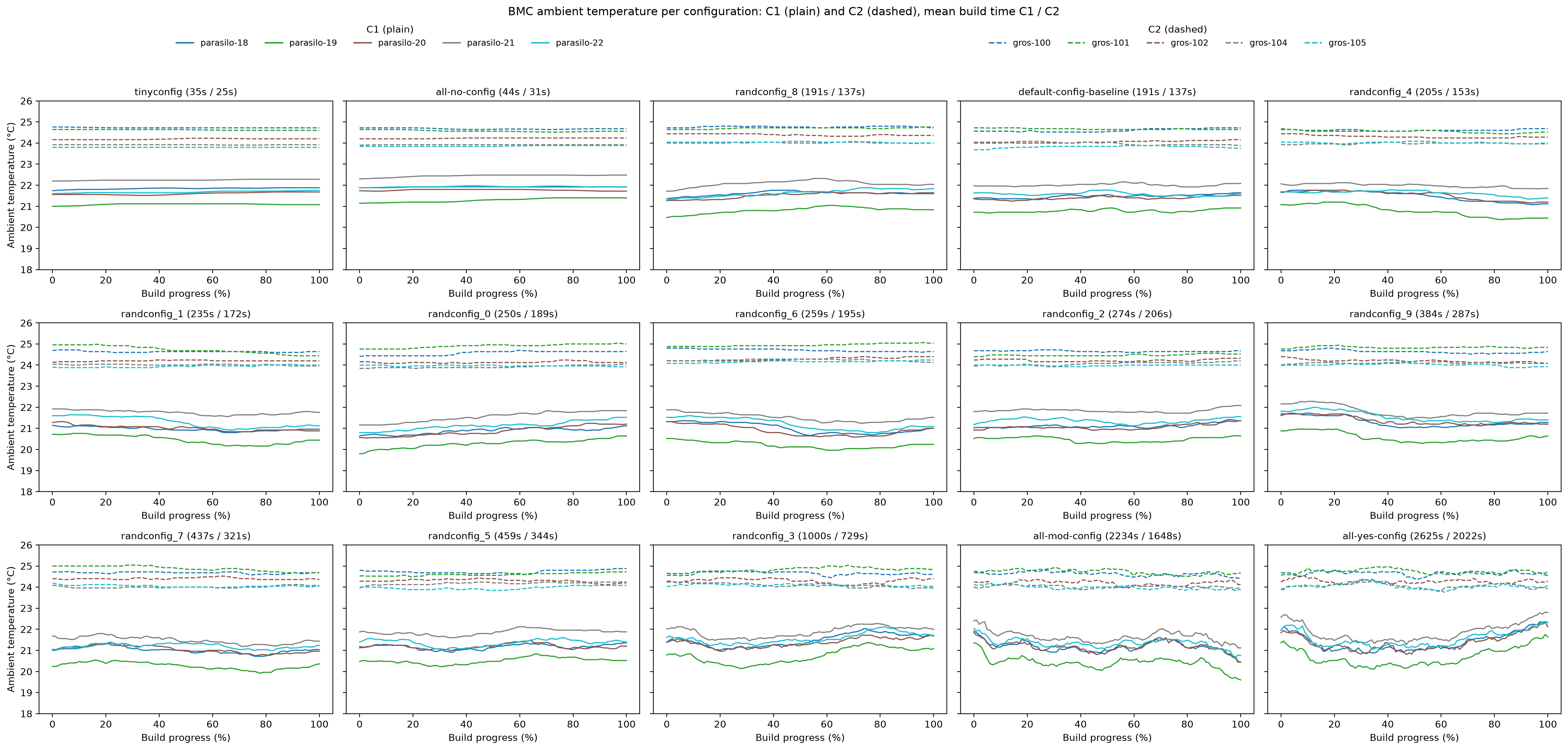}
  \caption{BMC ambient temperature during the builds, as a function of build progress, averaged over the 25 runs of each machine.
  \rennes is plain and \nancy dashed, with one colour per machine.
  Configurations are ordered by mean build time. Panel titles give the mean build time on \rennes (C1)/ \nancy (C2). All panels share the same y-axis.}
  \label{fig:app_temperature}
  \Description{BMC ambient temperature during the builds, as a function of build progress, averaged over the 25 runs of each machine (one colour per machine). Configurations are ordered by mean build time.}
\end{figure*}

\Cref{fig:app_temperature} shows the ambient temperature reported by the BMC of each machine during each build. For each run, the build window is rescaled to 0--100\% of build progress, and the profiles are averaged over the 25 runs of each machine. Configurations are ordered by mean build time and share the same y-axis.

\paragraph{\nancy is warmer but more stable.}
\nancy is always warmer than \rennes, by $2.4$ to $3.5$\,\textdegree C depending on the configuration (mean $24.3$ vs.\ $21.3$\,\textdegree C).
For every configuration, the coolest machine of \nancy is warmer than the warmest machine of \rennes.
\nancy is also more stable, in three ways:
\begin{itemize}
  \item \emph{Within a build}, the averaged profile varies by a median $0.2$\,\textdegree C on \nancy, against $0.6$\,\textdegree C on \rennes. 
  \item \emph{From run to run}, at the same build progress, the standard deviation is $0.6$\,\textdegree C on \nancy and $1.4$\,\textdegree C on \rennes.
  \item \emph{Across machines}, the machine means span $0.7$\,\textdegree C on \nancy ($24.0$--$24.7$\,\textdegree C) and $1.2$\,\textdegree C on \rennes ($20.6$--$21.9$\,\textdegree C).
\end{itemize}

\paragraph{Each machine keeps its own temperature.}
The order of the machines is the same in every configuration.
parasilo-19 is always the coolest machine of \rennes, and gros-100 and gros-101 are always the warmest of \nancy.
This systematic offset is likely due to the position of each machine in the rack and in the airflow of the room, rather than to the build.

\paragraph{Longer builds show more variation.}
The temperature variation within a build grows with its duration. Short builds such as \texttt{tinyconfig} and \texttt{allnoconfig} have flat profiles.
\allmodconfig and \allyesconfig, which last 30 to 45 minutes, show fluctuations of up to 1.8\,\textdegree C on \rennes.
This is expected: a longer build covers more samples and gives the room more time to change, so it captures more of the ambient variation.

\section{Build time per configuration and machine}
\label{app:build_time}

\begin{figure*}[t]
  \centering
  \includegraphics[width=\textwidth]{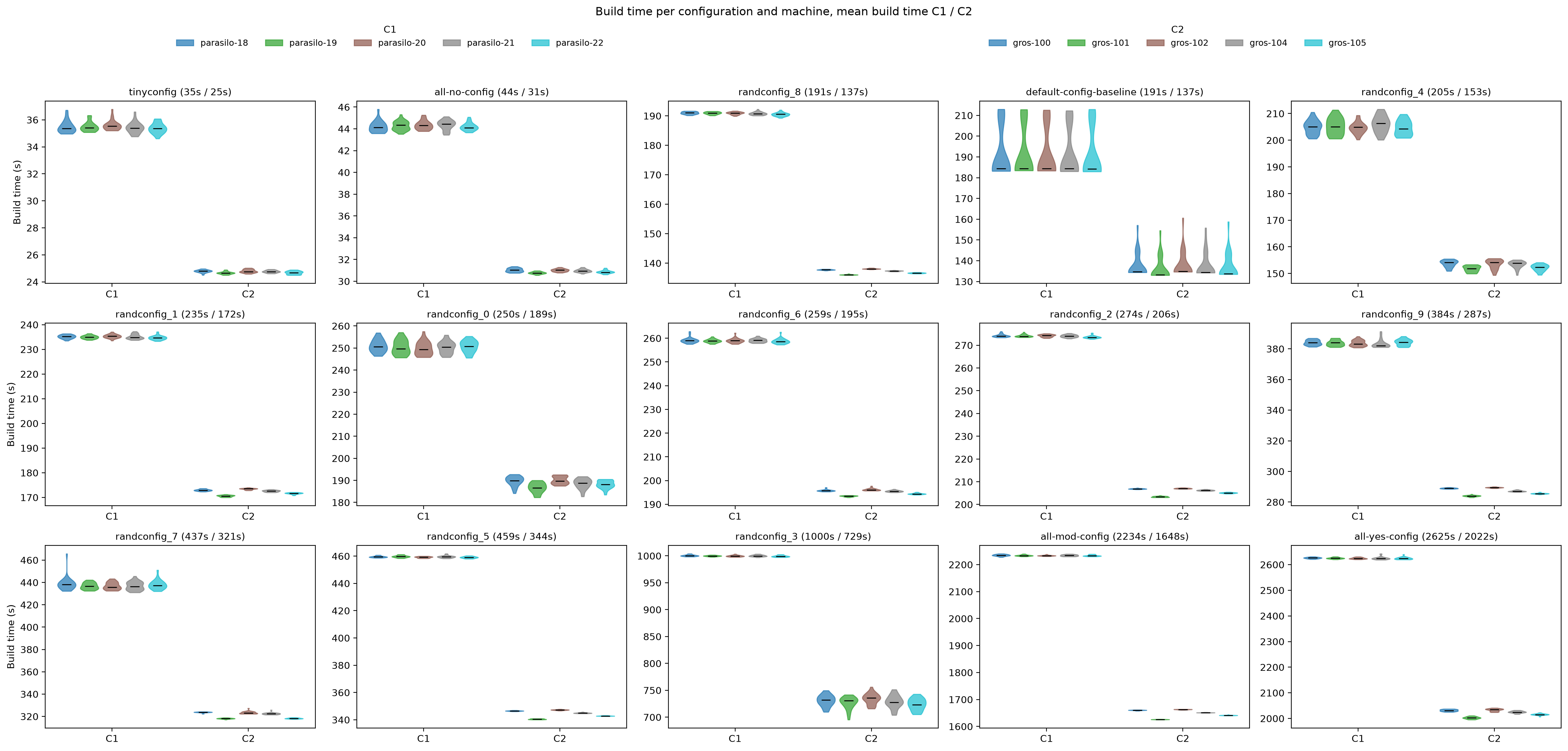}
  \caption{Build time of the 25 runs of each configuration on each machine (one violin per machine, black line at the median), on \rennes (C1, left of each panel) and \nancy (C2, right).
  Configurations are ordered by mean build time; panel titles give the mean build time on \rennes / \nancy.}
  \label{fig:app_build_time}
  \Description{Build time of the 25 runs of each configuration on each machine (one violin per machine, black line at the median), on both clusters.
  Configurations are ordered by mean build time.}
\end{figure*}

\Cref{fig:app_build_time} shows the build time of every run, per configuration and machine, on both clusters. Configurations are ordered by mean build time. Note that all panels have different y-axis.

\paragraph{The two clusters differ.}
\nancy builds every configuration faster than \rennes: its build time is $0.70$ to $0.77$ times that of \rennes (median $0.74$).
The ratio is almost the same for short builds (\texttt{tinyconfig}: 35\,s vs.\ 25\,s) and long builds (\texttt{allyesconfig}: 2{,}625\,s vs.\ 2{,}022\,s).
The two clusters therefore differ by a near-constant speed factor, consistent with their different processors.

\paragraph{Machines with the same hardware agree.}
Within a cluster, the five machines give almost the same build time.
The gap between the fastest and the slowest machine is a median $0.3\%$ of the build time on \rennes (at most $1.1\%$), and $1.7\%$ on \nancy (at most $2.2\%$).
The only exception is \texttt{default-config-baseline}, which has a few outlier runs. 
Build time is thus a reproducible quantity on a given cluster, whichever machine is used.



\end{document}